\documentclass[preprint,authoryear,12pt]{elsarticle}
\usepackage{graphicx}
\usepackage{amssymb}
\usepackage[ps2pdf,%
a4paper=true,%
breaklinks=true,%
colorlinks=true,%
pdfauthor={A. F. Zakharov et al.},%
pdftitle={Constraints on $R^n$ gravity from precession of orbits
of S2-like stars}%
]{hyperref}

\journal{Advances in Space Research}

\begin{document}
\begin{frontmatter}

\title{Constraints on $R^n$ gravity from precession of orbits
of S2-like stars: a case of a bulk distribution of mass}

\author[label1,label2,label2a,label2b,label2c]{A. F. Zakharov\corref{cor}}
\cortext[cor]{Corresponding author} \ead{zakharov@itep.ru}

\author[label3]{D. Borka}
\ead{dusborka@vinca.rs}

\author[label3]{V. Borka Jovanovi\'{c}}
\ead{vborka@vinca.rs}

\author[label4]{P. Jovanovi\'{c}}
\ead{pjovanovic@aob.rs}

\address[label1]{Institute of Theoretical and Experimental Physics,
 117218 Moscow, Russia}
\address[label2]{Bogoliubov Laboratory for Theoretical Physics, JINR,
141980 Dubna, Russia}
\address[label2a]{North Carolina Central University, 1801 Fayetteville
 Street, Durham, NC 27707, USA}
\address[label2b]{Institute for Computer Aided Design of RAS,     123056, Moscow, Russia}
 \address[label2c]{National Research Nuclear University (NRNU MEPHI), 115409,
   Moscow, Russia}

\address[label3]{Atomic Physics Laboratory (040), Vin\v{c}a Institute
of Nuclear Sciences, University of Belgrade, P.O. Box 522, 11001
Belgrade, Serbia}
\address[label4]{Astronomical Observatory, Volgina 7, 11060 Belgrade,
Serbia}

\begin{abstract}
Here we investigate possible applications of observed stellar orbits
around Galactic Center for constraining the R$^n$ gravity at
Galactic scales. For that purpose, we simulated orbits of S2-like
stars around the massive black hole at Galactic Center, and
study the constraints on the R$^n$ gravity which could be obtained by
the present and next generations of large telescopes. Our results
show that R$^n$ gravity affects the simulated orbits in the
qualitatively similar way as a bulk distribution of matter (including
a stellar cluster and dark matter distributions) in Newton's
gravity. In the cases where the density of extended mass is higher,
the maximum allowed value of parameter $\beta$ in R$^n$ gravity is
noticeably smaller, due to the fact that the both extended mass and
$R^n$ gravity cause the retrograde orbital precession.
\end{abstract}


\end{frontmatter}

\parindent=0.5 cm

\section{Introduction}
\label{Section 1} Dark matter (DM)  \citep{Zwicky_33} and Dark
Energy \citep{Turner_99}   problems are fundamental and difficult
for the conventional General Relativity approach for gravity,
\citep{Zakharov_09} see also the monograph by \cite{Weinberg_08} for
a more comprehensive review.

There is an opinion that an introduction of alternative theories of
gravity (including so-called $f(R)$ theories
\citep{capo02,capo03,capo06,carr04,capo07,capo11,mazu12}) could
after all give explanation of observational data without DM and DE
problems. However,  a proposed gravity theory has to explain not
only cosmological problems but many other observational data because
sometimes these theories do not have Newtonian limit for a weak
gravitational field case, so parameters of these theories have to be
very close to values which correspond to general relativity
\citep{Zakharov_06}. Earlier, constraints on $f(R)=R^n$ have been
obtained from an analysis of trajectories of bright stars near the
Galactic Center \citep{bork12,bork13}, assuming  a potential of bulk
distribution of matter is negligible in comparison with a potential
of a point like mass. In this paper we consider  modifications of
results due to a potential of a bulk distribution of matter assuming
that this potential is small in comparison with point like mass one.
For the standard GR approach these calculations have been done
\citep{Zakharov_07,Nucita_07}.

We would like to mention that not only trajectories of bright stars
but also probes such as LAGEOS and LARES
\citep{Ciufolini_04,Ciufolini_07} could provide important test for
alternative theories of gravity, see for instance constraints on the
Chern--Simons gravity \citep{Smith_08}.

\section{Method}
\label{Section 2} We simulated orbits of S2 star around Galactic
Center in the R$^n$ gravity potential, assuming a bulk distribution
of mass in the central regions of our Galaxy. The $R^n$ gravity
potential is given by \cite{capo06,capo07}:
\begin{equation}
\Phi \left( r \right) = -\frac{GM_{BH}}{2r}\left[ {1 + \left(
{\frac{r}{r_c}} \right)^\beta} \right], \label{rngpot}
\end{equation}%
where $r_c$ is an arbitrary parameter, depending on the typical
scale of the considered system and $\beta$ is a universal constant
depending on $n$ \citep{capo06,Zakharov_06}
\begin{equation}
\beta=\frac{12n^2-7n -1
-\sqrt{36n^4+12n^3-83n^2+50n+1}}{6n^2-4n+2}~.
\end{equation}

We use two component model for a potential in the central region of
our Galaxy
which is constituted by the central black hole of mass $M_{BH}= 4.3
\times10^6 M_\odot$ \citep{gill09} and an extended distribution of
matter with total mass $M_{ext}(r)$ (including a stellar cluster and
dark matter) contained within some radius $r$. For the density
distribution of extended matter we adopted the following broken
power law
 proposed by \cite{genz03}:
\begin{equation}
\rho(r)=\rho_0\left( \frac{r}{r_0}\right) ^{-\alpha },\;\alpha
=\left\{
\begin{array}{ll}
2.0\pm 0.1, & r\geq r_0 \\
1.4\pm 0.1, & r < r_0
\end{array}
\right.
\label{rho}
\end{equation}%
where $\rho_0=1.2\times 10^{6}\, M_{\odot }\cdot\mathrm{pc}^{-3}$ and
$r_0=10^{\prime\prime}$.
This leads to the following expression for the extended mass
distribution:
\begin{equation}
M_{ext}(r)=\frac{4\pi\rho_0 r_0^\alpha}{3-\alpha}r^{3-\alpha}.
\label{extmass}
\end{equation}%

\begin{figure*}[ht!]
\centering
\includegraphics[width=0.49\textwidth]{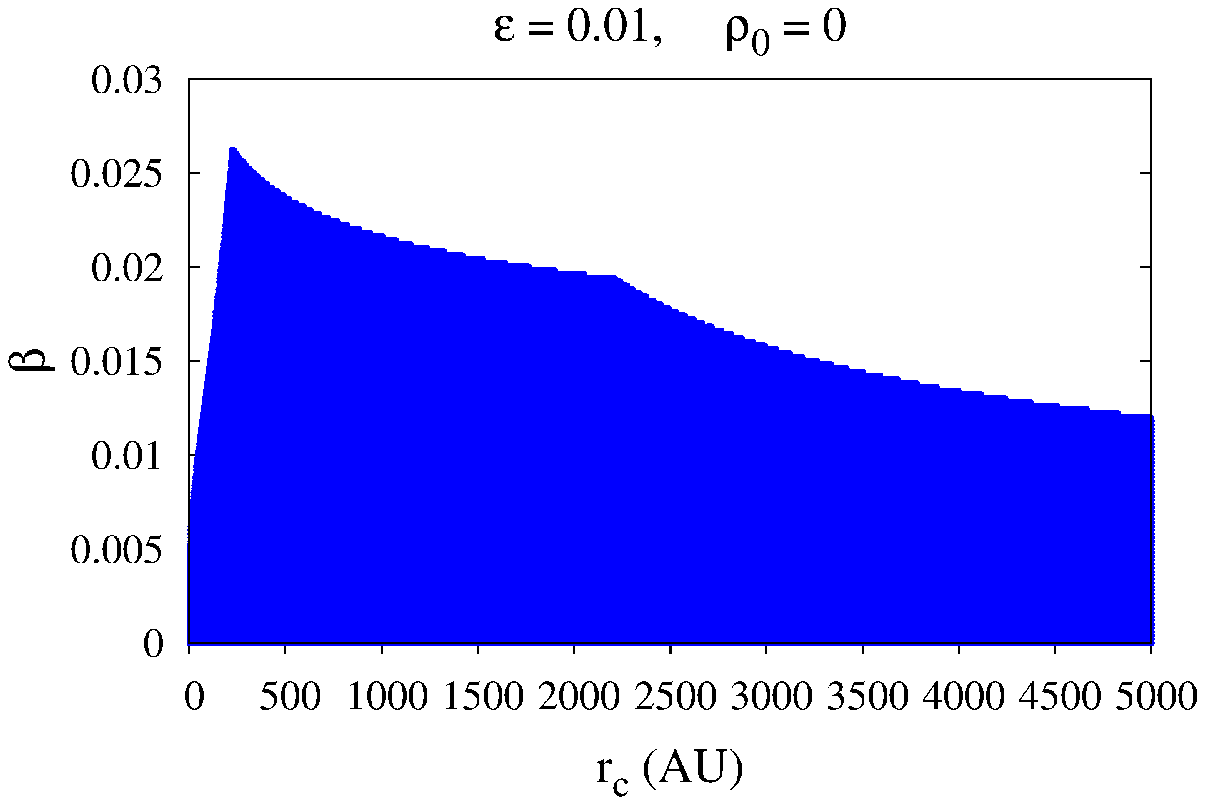}
\hfill
\includegraphics[width=0.49\textwidth]{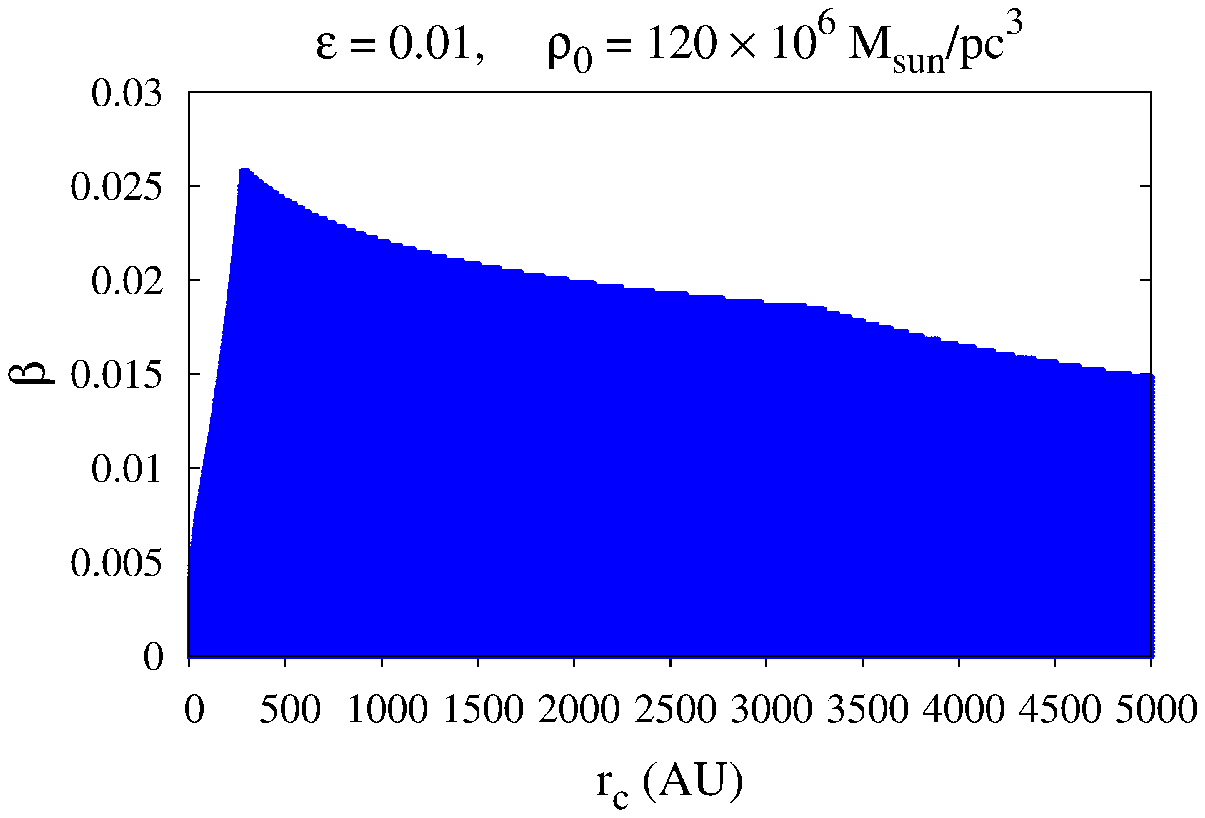}
\caption{Parameter space for $R^n$ gravity with different
contributions of extended mass under the constraint that, during one
orbital period, S2-star orbits in $R^n$ gravity differ less than 10
mas ($\varepsilon=0''.01$) from its Keplerian orbit. The assumed
values for mass density constant $\rho_0$ from Eq. (\ref{rho}) are
denoted in the title of each panel.}
\label{fig01}
\end{figure*}

\begin{figure*}[ht!]
\centering
\includegraphics[width=0.49\textwidth]{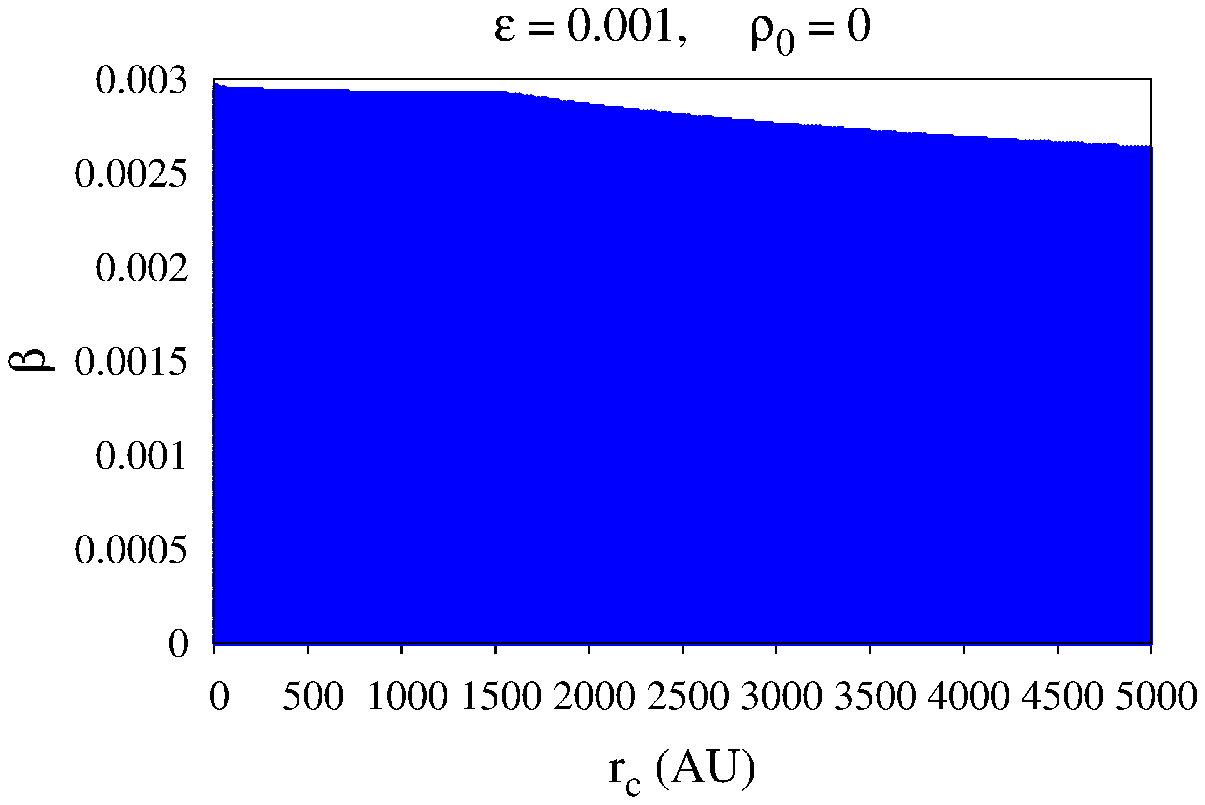}
\hfill
\includegraphics[width=0.49\textwidth]{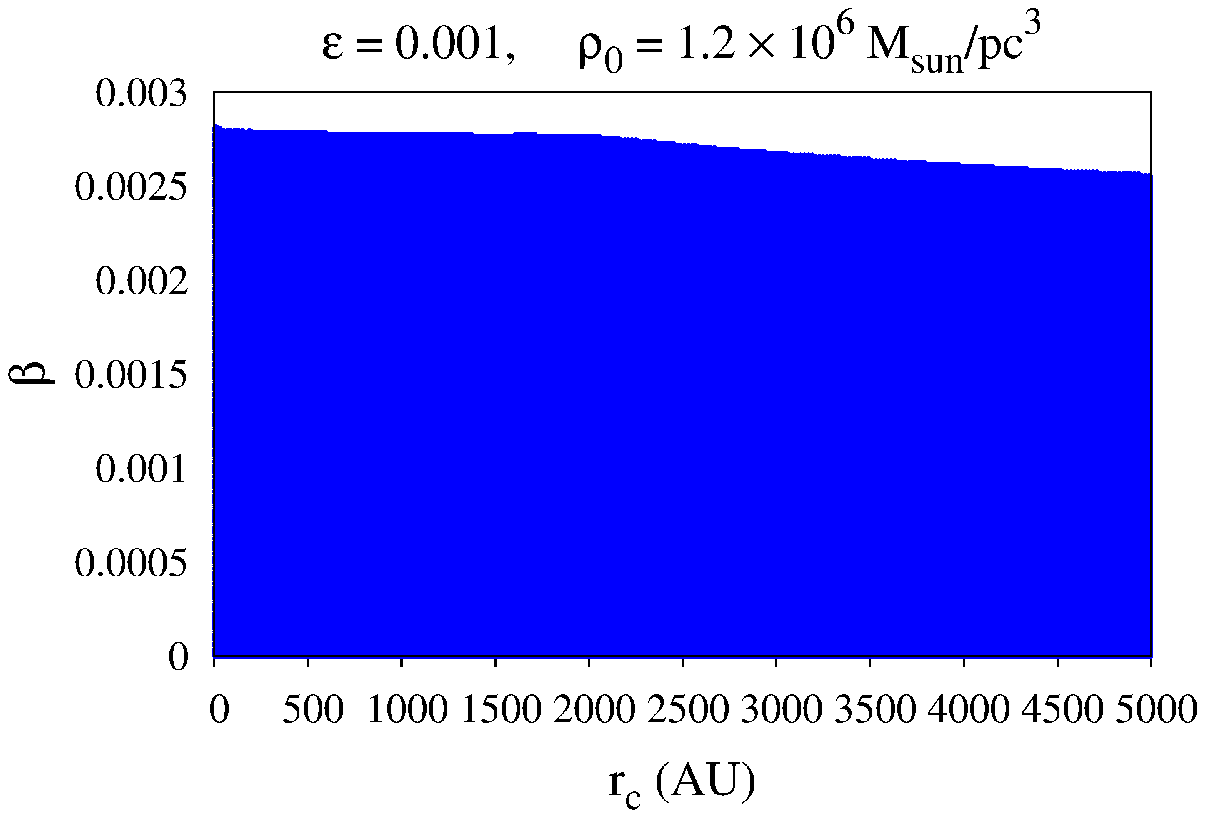}\\
\vspace{0.4cm}
\includegraphics[width=0.49\textwidth]{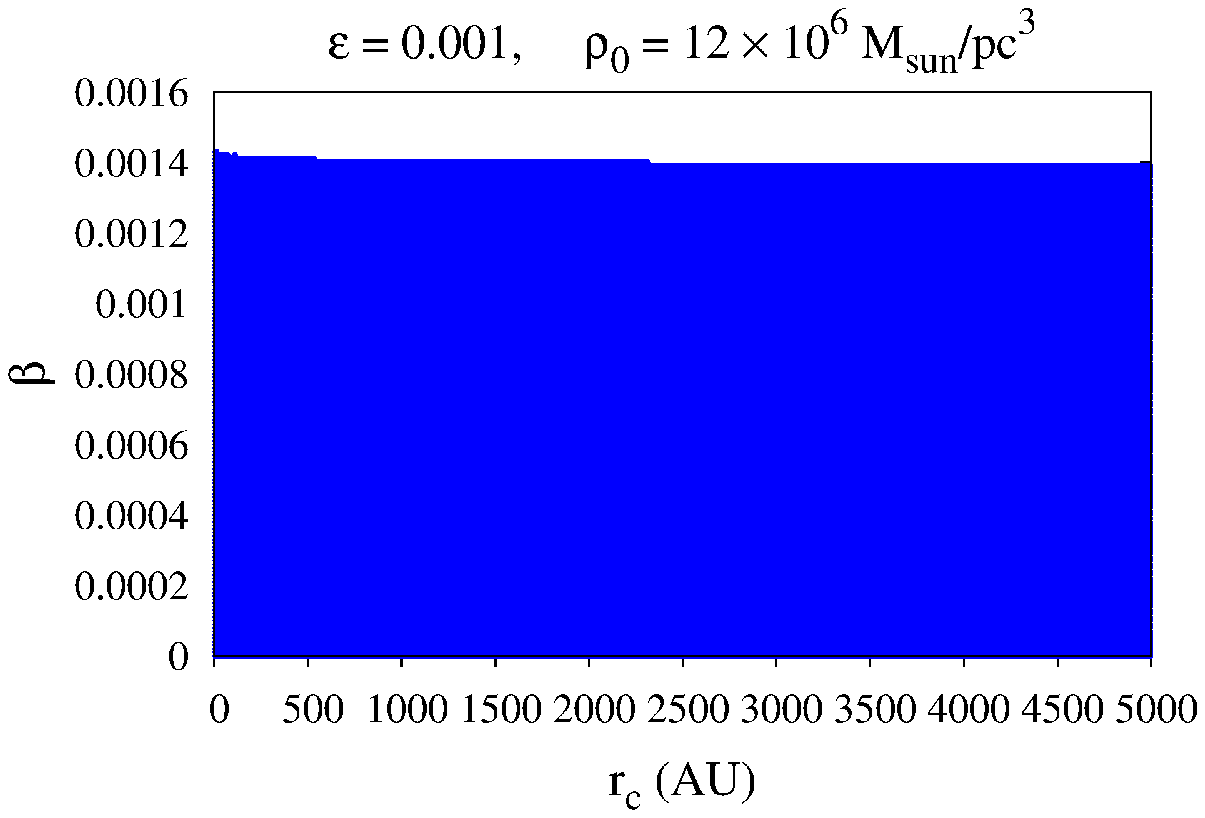} 
\hfill
\includegraphics[width=0.49\textwidth]{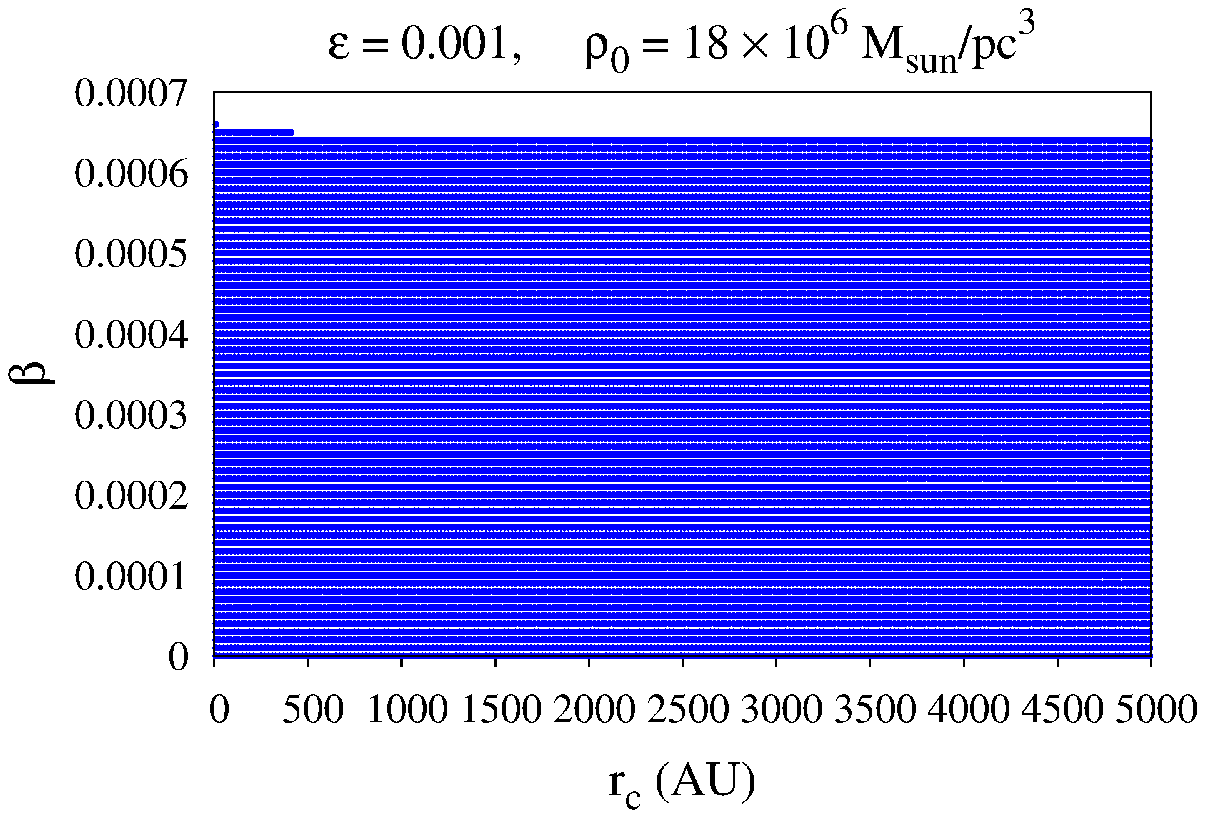}
\caption{The same as in Fig. \ref{fig01}, but for 10 times
higher astrometric precision of 1 mas ($\varepsilon=0''.001$).}
\label{fig02}
\end{figure*}

\begin{figure*}[ht!]
\centering
\includegraphics[width=0.49\textwidth]{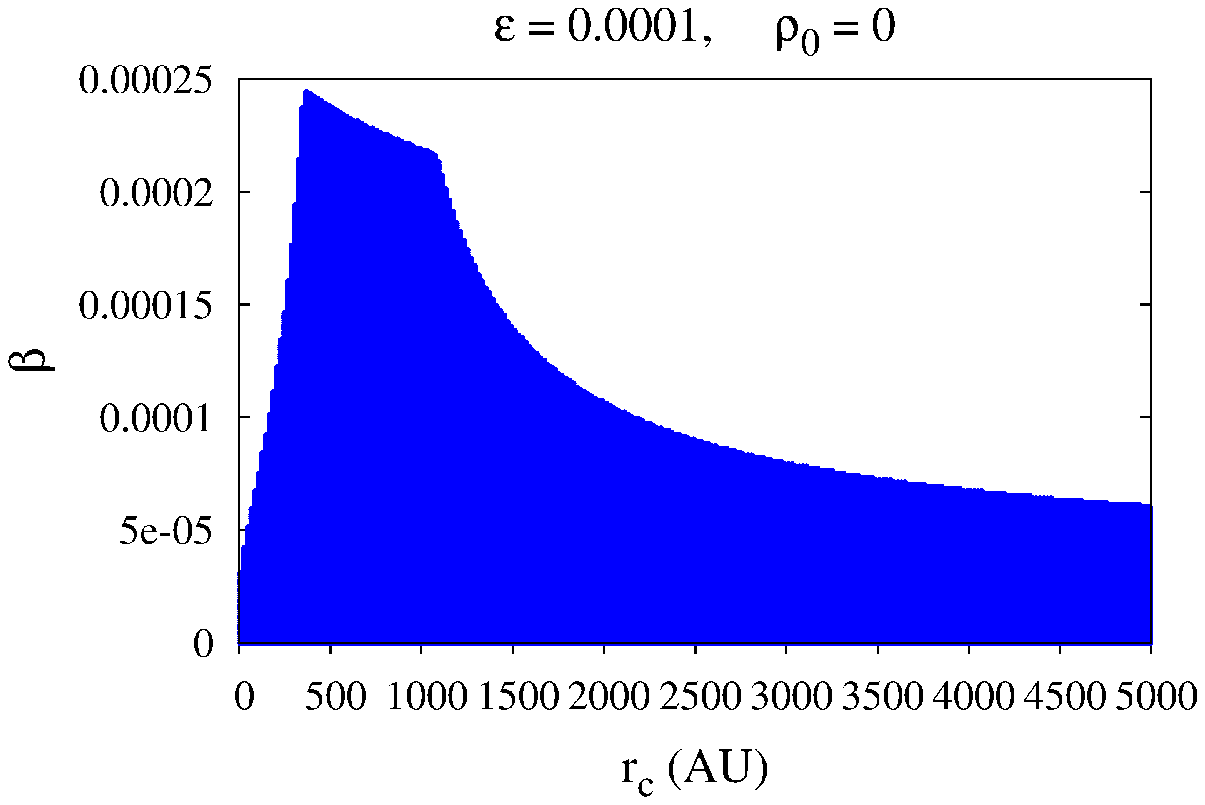}
\hfill
\includegraphics[width=0.49\textwidth]{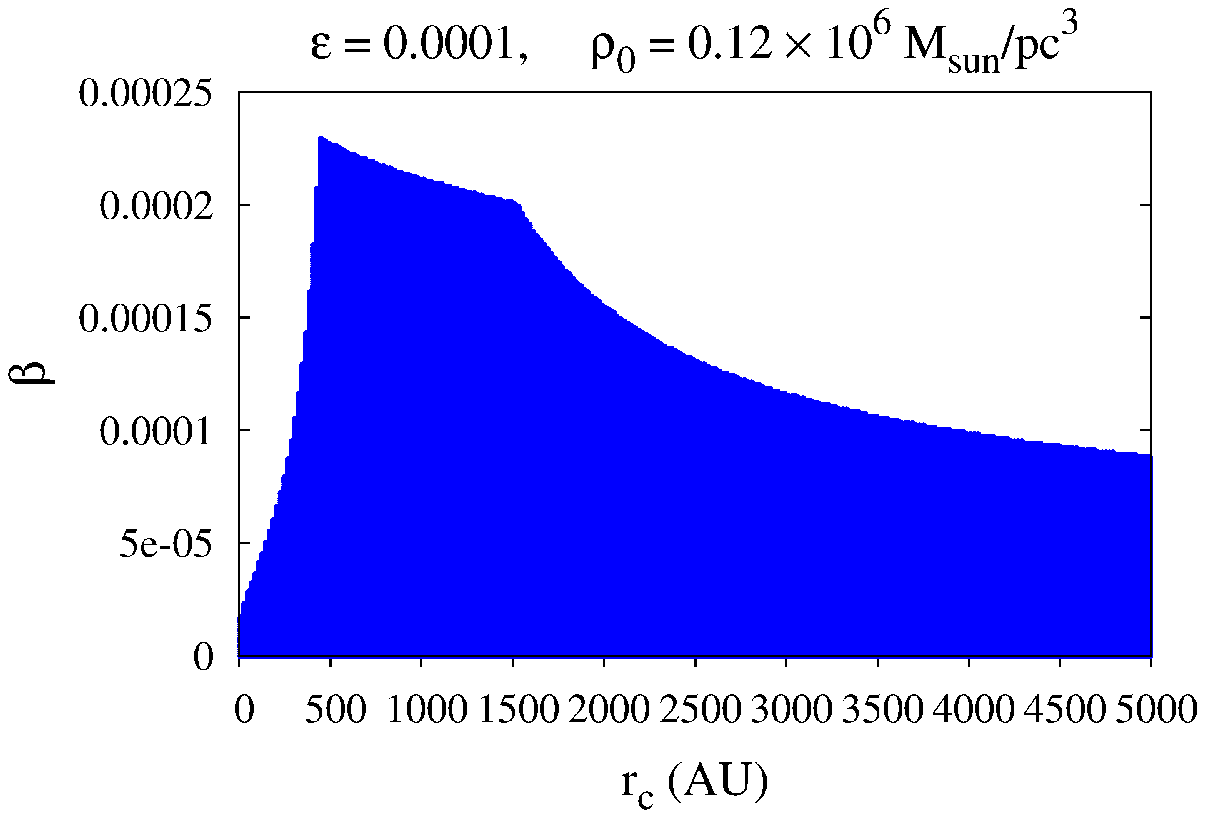} \\
\vspace{0.4cm}
\includegraphics[width=0.49\textwidth]{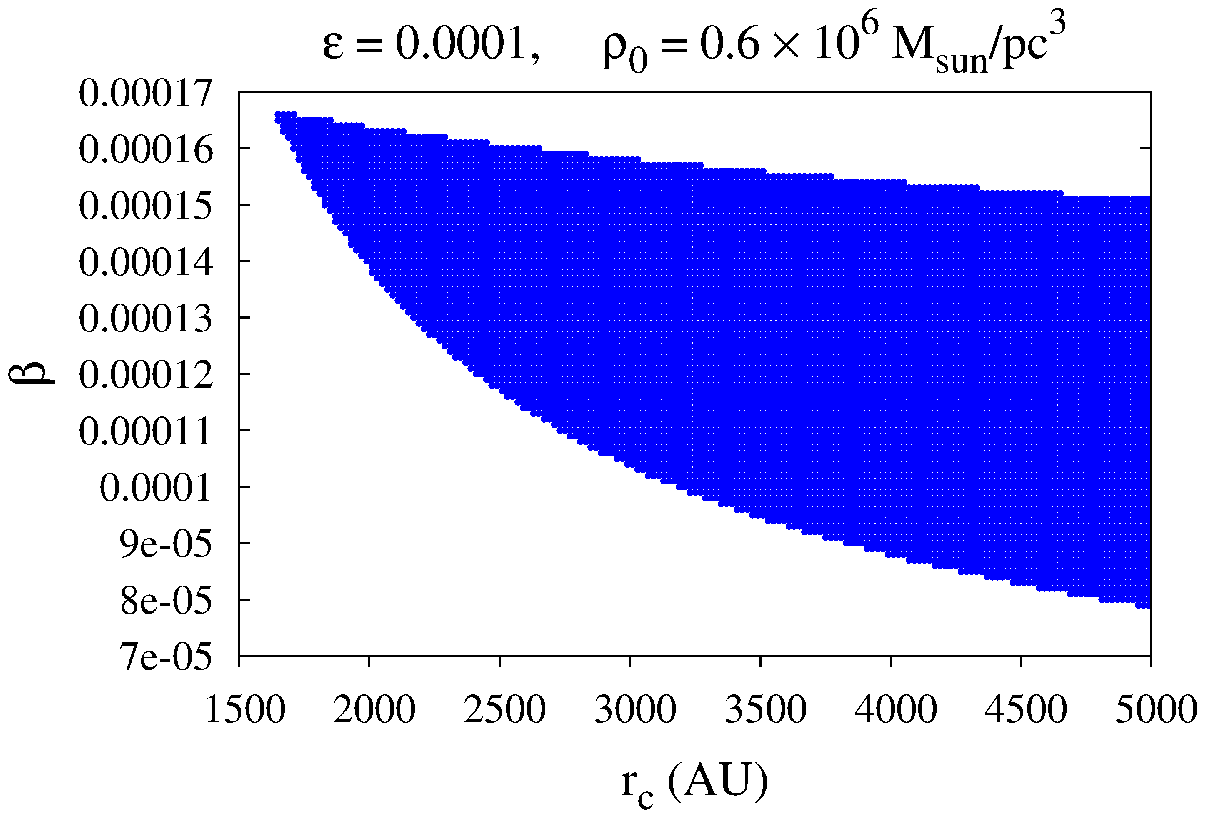}
\caption{The same as in Fig. \ref{fig01}, but for 100 times
higher astrometric precision of 0.1 mas ($\varepsilon=0''.0001$).}
\label{fig03}
\end{figure*}

\begin{figure*}[ht!]
\centering
\includegraphics[width=0.49\textwidth]{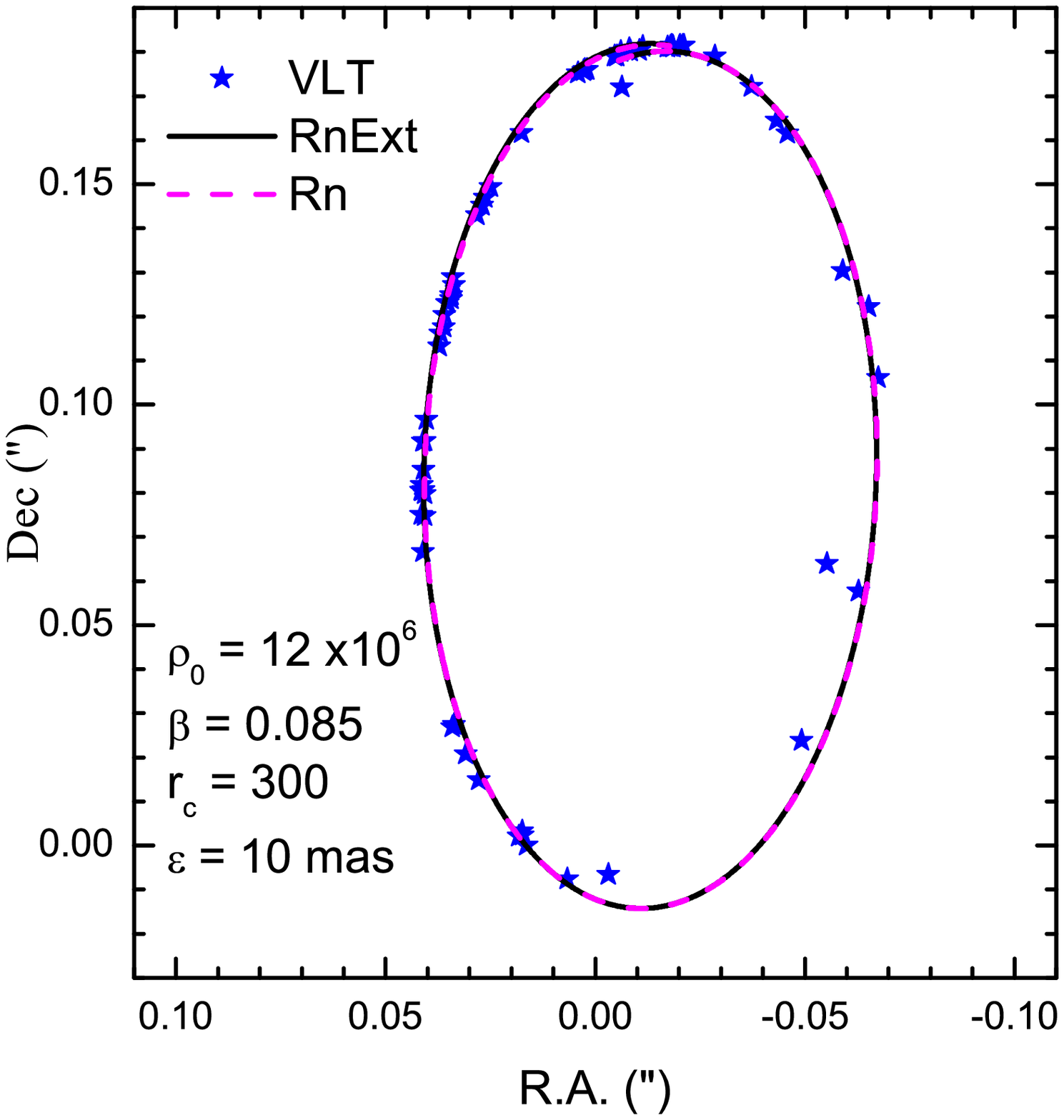}
\hfill
\includegraphics[width=0.49\textwidth]{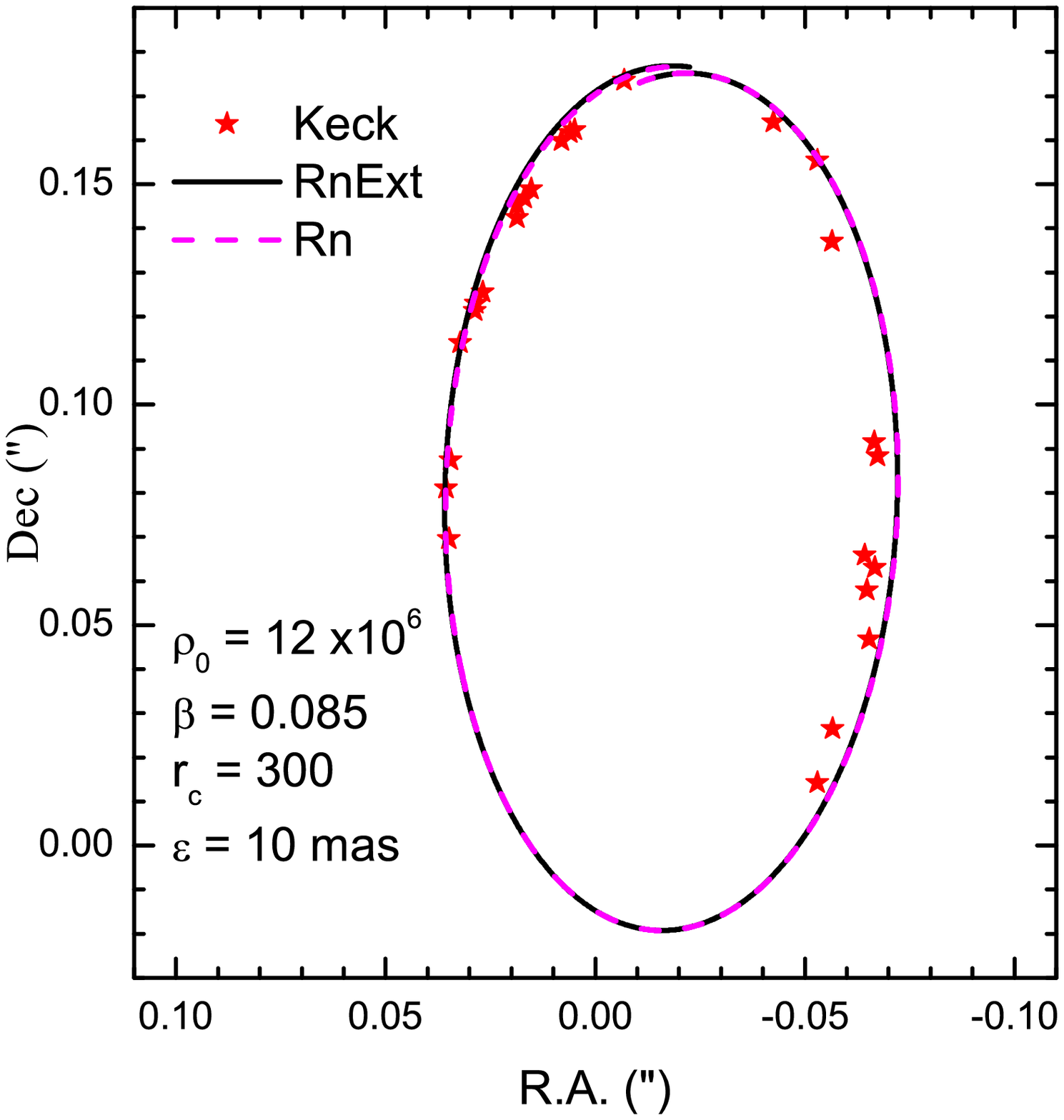} \\
\caption{Comparison between the simulated orbits of S2 star in the
R$^n$ gravity potential with (solid line) and without (dashed line)
contribution of extended mass, and NTT/VLT (left) and Keck
(right) astrometric observations. The parameters of R$^n$ gravity are
obtained by fitting the simulated orbits in total potential to the
observations, assuming astrometric accuracy of 10 mas and the
density constant $\rho_0=12\times 10^{6}\,
M_{\odot }\cdot\mathrm{pc}^{-3}$.}
\label{fig04}
\end{figure*}


{The corresponding potential for extended distribution of matter is
then:
\begin{equation}
\Phi_{ext}(r)=-G\int_r^{r_\infty}\frac{M_{ext}(r^\prime)}{r^{\prime
2}}dr^\prime = \frac{{ - 4\pi {\rho _0}R_0^\alpha G}}{{\left( {3 -
\alpha } \right) \left( {2 - \alpha } \right)}}\left( {{r_\infty}^{2
- \alpha } - {r^{2 - \alpha }}} \right), \label{extpot}
\end{equation}%
where $r_\infty$ is the outer radius for extended distribution of
matter which is enclosed within the orbit of S2 star. The total
gravitational potential for the two component  model can be
evaluated as a sum of $R^n$ potential for central object with mass
$M_{BH}$ and potential for extended matter with mass $M_{ext}(r)$:
\begin{equation}
\Phi_{total}(r)=\Phi(r)+\Phi_{ext}(r).
 \label{pot_total}
\end{equation}%
 Thus, the simulated orbits of S2 star could be obtained by
numerical integration of the following differential equations of
motion in the total gravitational potential:
\begin{equation}
\mathbf{\dot{r}}=\mathbf{v},\hspace*{0.5cm}
\mathbf{\ddot{r}}=-\nabla\left(\Phi_{total}\left(\mathbf{r}\right))\right).
\label{2body}
\end{equation}
}

\section{Results}
\label{Section 3}

In Figs. \ref{fig01}--\ref{fig03} we present the parameter space for
$R^n$ gravity with different contributions of extended mass for
which the discrepancies between the orbits of S2-star in $R^n$
gravity and its Keplerian orbit during one orbital period are less
than an assumed astrometric precision. As it can be seen from Fig.
\ref{fig01}, it is very difficult to detect the contribution of
extended mass with the astrometric precision of 10 mas, which was
the actual limit during the first part of the observational period
of S2 star, some 10--15 years ago. The blue area of parameter space
changing very little with variations of $\rho_0$. However, with the
current astrometric limit reaching less than 1 mas, this
contribution significantly constrains the maximum allowed value
$\beta$ (see Fig. \ref{fig02}). In the cases where the density of
extended mass is higher, the maximum allowed value of $\beta$ is
noticeably smaller, due to the fact that the both extended mass and
$R^n$ gravity have the similar effect on S2 star orbit, i.e. they
both cause the retrograde orbital precession. However, the
astrometric limit is constantly improving and in the future it will
be possible to measure the stellar positions with much better
accuracy of $\sim 10\ \mu$as \citep{gill10}. The parameter space for
currently unreachable accuracy of 0.1 mas is presented in Fig.
\ref{fig03}, from which one can see that even a small amount of
extended mass would practically exclude the $R^n$ term from the
total gravity potential. Even more, we found that in such a case
expression (\ref{rho}) would result with overestimated amount of
extended mass at Galactic center, and therefore, we assumed 2 and 10
times smaller densities $\rho_0$. Besides, from Figs.
\ref{fig01}--\ref{fig03} it is obvious that both astrometric
precision and amount of extended mass have significant influence on
the value of $r_c$ for which maximum $\beta$ is expected. For
example, in the top right panel of Fig. \ref{fig01} the maximum
$\beta \approx 0.027$ is expected for $r_c \approx 250$ AU, while in
the top right panel of Fig. \ref{fig03} the maximum $\beta \approx
0.00023$ is expected for $r_c \approx 450$ AU.

Several comparisons of the simulated S2 star orbits in the R$^n$
gravity potential with and without contribution of extended mass
with NTT/VLT and Keck astrometric observations are given in Fig.
\ref{fig04}, assuming the astrometric accuracy of 10 mas. In this
case, influence of extended mass for $\rho_0=1.2\times10^6 \,
M_{\odot }\cdot\mathrm{pc}^{-3}$ \citep{genz03} is almost
negligible, and hence it cannot explain the observed precession of
S2 star orbit. Therefore, in order to explain the observed
precession, either a higher value of $\beta$ in $R^n$ gravity
potential \citep[see e.g.][]{bork12}, or much higher density of
extended mass are necessary (as it is the case in Fig.
\ref{fig04}
).

\clearpage

\section{Conclusions}

In this paper we analyze
stellar orbits around Galactic Center in the R$^n$ gravity with
extended mass distribution. Our results show
that R$^n$ gravity with extended mass distribution could
significantly affect the simulated orbits. Both $R^n$ gravity and
extended mass distribution give retrograde direction of the
precession of the S2 orbit. In the cases if the density of
extended mass is higher, the maximum value of $\beta$ which is
consistent with observations in R$^n$ gravity is noticeably smaller.
We confirmed that the R$^n$ gravity parameter $\beta$ must be very
close to those corresponding to the Newtonian limit of the theory.
When parameter $\beta$ is vanishing, we recover the value of the
Keplerian orbit for S2 star. When we take into account extended mass
distribution, parameter $\beta$ is less than in case without
extended mass distribution and faster approaching to zero. Even
more, one can see that relatively small amount of extended mass
would practically exclude the $R^n$ term from the total gravity
potential.

We can conclude that both effects, additional term in $R^n$
gravity and extended mass distribution, produce a retrograde shift,
that results in rosette shaped orbits. Also, we can conclude that
both astrometric precision and extended mass distribution
have significant influence on the value of $r_c$ for which maximum
$\beta$ could be expected.

Although both observational sets (NTT/VLT and Keck) indicate that
the orbit of S2 star might not be closed, the current astrometric
limit is not sufficient to unambiguously confirm such a claim.
However, the astrometric accuracy is constantly improving from
around 10 mas during the first part of the observational period,
currently reaching less than 1 mas.

\paragraph{Acknowledgments}
D. B., V. B. J. and P. J. acknowledge support of the Ministry of
Education, Science and Technological Development of the Republic of
Serbia through the project 176003 ''Gravitation and the large scale
structure of the Universe''. A. F. Z. acknowledges a partial support
of the NSF (HRD-0833184) and NASA (NNX09AV07A) grants at NCCU
(Durham, NC, USA) and RFBR 14-02-00754a at ICAD of RAS (Moscow).
Authors thank anonymous referees for their useful critical remarks.



\end{document}